%% file: IEEE.ICC.2026_Analysis_of_NTN_by_SNC.tex
\documentclass[conference,twocolumn,10pt]{IEEEtran}
\usepackage[utf8]{inputenc}
\usepackage{graphicx}
\usepackage{amssymb}
\usepackage{mathtools}
\usepackage{dsfont}
\usepackage{cite}
\usepackage{stfloats}
\usepackage{subfigure}
\usepackage{psfrag}
\usepackage[mathscr]{euscript}
\usepackage{acronym}  
\usepackage{amsmath}
\usepackage{algorithm}
\usepackage[noend]{algpseudocode}
\usepackage{booktabs}
\usepackage{bbm}
\usepackage{dsfont}
\usepackage{verbatim} 
\usepackage{soul}

\usepackage{times}
\usepackage{xcolor}
\usepackage{textcomp}
\usepackage{gensymb}
\usepackage{amsthm}

\makeatletter
\renewcommand{\maketag@@@}[1]{\hbox{\m@th\normalsize\normalfont#1}}%

\makeatother

\makeatletter
\def\BState{\State\hskip-\ALG@thistlm}
\makeatother

\usepackage{float}
\usepackage{balance}
\input{SupportDocuments/aconym}
\input{SupportDocuments/defmetric}

\newcommand{\paperTitle}{
Analysis of SINR Coverage in LEO Satellite Networks through Spatial Network Calculus
}

\begin{document}

\title{\paperTitle}
\author{
    \IEEEauthorblockN{Yuting~Tang\IEEEauthorrefmark{1}, Yufan~He\IEEEauthorrefmark{1}, Yi~Zhong\IEEEauthorrefmark{2}, 
  Xijun~Wang\IEEEauthorrefmark{3},
  Tony Q. S. Quek\IEEEauthorrefmark{4},
  and Howard~H.~Yang\IEEEauthorrefmark{1}}
  \IEEEauthorblockA{\IEEEauthorrefmark{1}%
                    ZJU-UIUC Institute, Zhejiang University, Haining, China}
                    
  \IEEEauthorblockA{\IEEEauthorrefmark{2}%
                    School of Electronic Information and Communications, Huazhong University of Science and Technology, Wuhan, China}

  \IEEEauthorblockA{\IEEEauthorrefmark{3}%
                    School of Electronics and Information Technology, Sun Yat-sen University, Guangzhou, China}        
  \IEEEauthorblockA{\IEEEauthorrefmark{4}%
                    Information Systems Technology and Design, Singapore University of Technology and Design, Singapore}  
}

\maketitle
\acresetall
\thispagestyle{empty}

\begin{abstract}
We introduce a new analytical framework, developed based on the spatial network calculus, for performance assessment of Low Earth Orbit (LEO) satellite networks. 
Specifically, we model the satellites' spatial positions as a strong ball-regulated point process on the sphere.
Under this model, proximal points in space exhibit a locally repulsive property, reflecting the fact that intersatellite links are protected by a safety distance and would not be arbitrarily close. 
Subsequently, we derive analytical lower bounds on the conditional coverage probabilities under Nakagami-\textit{m} and Rayleigh fading, respectively.
These expressions have a low computational complexity, enabling efficient numerical evaluations. 
We validate the effectiveness of our theoretical model by contrasting the coverage probability obtained from our analysis with that estimated from a Starlink constellation. 
The results show that our analysis provides a tight lower bound on the actual value and, surprisingly, matches the empirical simulations almost perfectly with a 1 dB shift. 
This demonstrates our framework as an appropriate theoretical model for LEO satellite networks. 
\end{abstract}


\acresetall

\section{Introduction}\label{sec:intro}
How to efficiently evaluate the link performance in a satellite network?
A line of studies \cite{Talgat2021StocGA,Wang2022UldenseLEO,AlHourani2021AnAA,Okati2020DownlinkCA,Okati2021NonhomogeneousSG} suggested using stochastic geometry.
Indeed, over the past decade, stochastic geometry has proven itself an effective theoretical tool for the design and analysis of terrestrial networks \cite{Haenggi2009steo}.
By modeling the locations of base stations as spatial point processes (most often Poisson), one could obtain elegant mathematical expressions for key performance metrics such as coverage probability \cite{Andrews2011Coverage}, throughput \cite{Zhang2015Throughput}, and delay \cite{Haenggi2013delay}, enabling a holistic assessment of scaling effects as well as optimization of the network.
Inspired by this, a natural approach is to extend the framework from terrestrial to non-terrestrial settings, which resulted in several analytical models for satellite networks, developed based on homogeneous Poisson point processes (HPPPs) \cite{AlHourani2021AnAA}, Binomial point processes (BPPs) \cite{Okati2020DownlinkCA}, and non-homogeneous Poisson point processes (NPPPs) \cite{Okati2021NonhomogeneousSG}.
However, these are two critical issues with these models. 
One, neither PPP nor BPP can rule out the possibility that two points are located arbitrarily close in proximity, which does not align with reality -- satellites simply do not collide in the sky.
The other, even the setback of these theoretical models is acceptable; due to the spherical structure of satellite networks, the resultant analytical expressions often involve layers of integrals, where numerical evaluation can take as long as running an empirical simulation directly, rendering the analysis ineffective.
In light of these challenges, we leverage the spatial network calculus \cite{Feng2024SNC}, a recently developed network model, to establish a more realistic network model and, concurrently, obtain less complex yet (very) accurate analytical results. 

\subsection{Prior Art}
Several existing works have developed mathematical models to characterize the spatial topology of satellite networks based on BPP \cite{Okati2020DownlinkCA}, HPPP \cite{AlHourani2021AnAA}, and NPPP \cite{Okati2021NonhomogeneousSG}, accounting for effects such as multi-layered constellation \cite{Talgat2021StocGA} and/or coordinated beamforming \cite{Kim2023CoverageAO}.
These analytical results enable one to quantify various performance metrics, ranging from coverage probability \cite{Park2023ATA}, spectral efficiency \cite{kim2024spectrumsharing}, and positioning accuracy \cite{WanKisYan:25TAES} and age of information \cite{LuYanPap:23Globecom}.
Recognizing the fact that in practice, nearby satellites will be separated by a guarded distance, 
\cite{Zhang2025PAHPP} incorporates inter-satellite safety distances into the modeling of multi-orbit. These multi-satellite communication systems significantly enhanced the capability for realistic satellite analysis. However, this model is limited to maintaining safety distances for satellites within the same orbit (i.e., intra-orbit), neglecting potential collisions between satellites on intersecting orbits (i.e., inter-orbit).

\subsection{Contributions}
Based on the spatial network calculus \cite{Feng2024SNC}, \cite{Zhong2025SNC}, we develop new theoretical models and analytical results for the coverage probability of LEO satellite networks. 
The main contributions are summarized below.

\begin{itemize}
    \item We establish the notion of strong ball-regulated point processes on a spherical surface. 
    Under such a model, the points exhibit a locally repulsive pattern, prohibiting any two points from locating arbitrarily close. 
    Hence, it can serve as an appropriate spatial model for satellite networks. 

    \item We derive analytical expressions for the (worst-case) lower bound to the satellite network's coverage probability, under Nakagami-\textit{m} fading and Rayleigh fading, respectively. 
    The theoretical results have low computational complexity and can be efficiently evaluated numerically. 

    \item 
    We validate the effectiveness of our analytical framework by comparing the theoretical lower bounds against simulations based on Starlink constellation data. 
    Surprisingly, we find that the simulation and analysis results amid an almost perfect match via a 1 dB translation, which corroborates the efficacy of our method.  
\end{itemize}

\section{System Model}\label{sec:sysmod}

\begin{figure}[t!]
  \centering{}
    {\includegraphics[width=0.9\columnwidth]{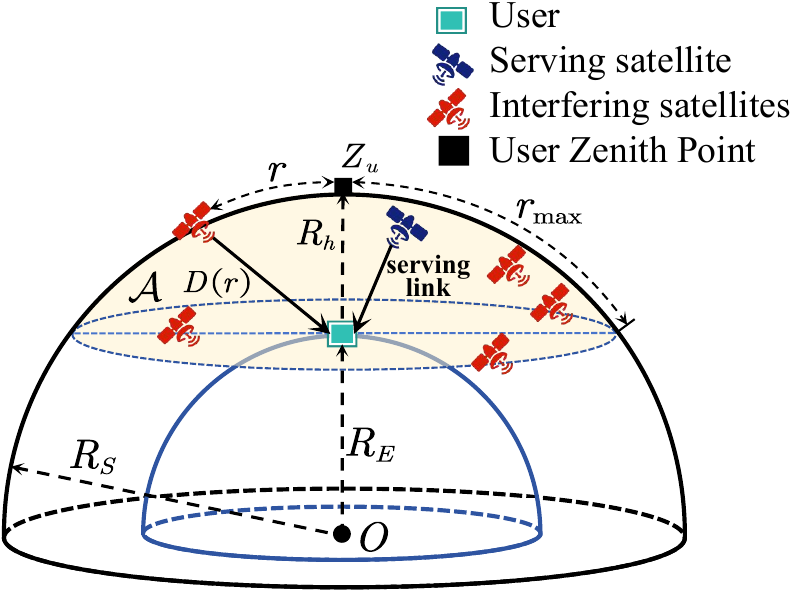}}
  \caption{A snapshot of the satellite downlink network model. 
  A user is located on the Earth's surface. The blue icon represents the serving satellite, while the red icons denote the interfering satellites within the shaded visible cap. 
 }
  \label{fig:satellite_model}
  \vspace{-10pt}
\end{figure}

\subsection{Network Configuration}
We consider the downlink of a LEO satellite network, as depicted in Fig.~\ref{fig:satellite_model}. The satellites are deployed on a sky sphere $\mathcal{S}_{R_\mathrm{S}}$ of radius $R_\mathrm{S}$, concentric with the Earth. 
We denote by $\mathcal{S}_{R_\mathrm{E}}$ the surface of the Earth, modeled as another sphere of radius $R_{ \mathrm{E} }$.
As such, the altitude of each satellite is $R_\mathrm{h} = R_\mathrm{S}-R_\mathrm{E}$.
We model the satellite locations as a stationary, isotropic point process (which is non-Poisson, as we will elaborate in the next section), denoted by $\Phi$.
Without loss of generality, we place the user at the North Pole, with coordinates $ u = (0, 0, R_\mathrm{E})$.
We denote by $\mathcal{B} \subset \mathcal{S}_{R_\mathrm{S}}$ the visible region from the user's perspective, defined as the spherical dome determined by the intersection of the sky sphere with the tangent plane to the Earth at $u$.
Note that only satellites located within the visible spherical cap $\mathcal{B}$ are observable to the user. 

We assume all satellites transmit at a constant power $P_\mathrm{tx}$ over a shared frequency band. 
In this network, the radio signal propagation is affected by two major factors: large-scale path loss and small-scale channel fading.
Specifically, we use a power-law model to characterize the path loss and assume that the channel fading follows the Nakagami-\textit{m} distribution \cite{Okati2021NonhomogeneousSG}. Moreover, we assume the channel fading of all links is spatially and temporally independent \cite{haenggi2012HCPP}.

We model the antenna gain of each satellite using a two-level directional gain pattern \cite{Park2023ATA}, comprised of a main-lobe gain $G_t^M$ and a side-lobe gain $G_t^S$. For the typical user, the serving satellite aligns its main lobe ($G_t^M$) with the user, while all interfering satellites are assumed to point a side lobe ($G_t^S$) towards the user \cite{Park2023ATA}.
As such, the effective antenna gain $G_i$ between a transmitting satellite $x_i$ and the user is given by
\begin{align}
    G_i=
    \begin{cases}
        G_t^M G_r \frac{c^2}{(4\pi f_c)^2}, & \text{$x_i$ is the serving satellite;} \\
        G_t^S G_r \frac{c^2}{(4\pi f_c)^2}, & \text{otherwise,}
    \end{cases}
\end{align}
where $G_r$ is the user's receiver gain, $c$ is the speed of light, and $f_c$ is the carrier frequency.

\subsection{Performance Metric}
At the user's standpoint, given the position of a visible satellite at $x_i \in \Phi$, we denote by $r_i = d_{\mathcal{S}} ( Z_u, x_i ) $ the \textit{spherical distance} (measured on the sky sphere $\mathcal{S}_{R_{ \mathrm{S} }}$) between the user's zenith point $Z_u$ and this satellite.   
Due to the visibility constraint, we have $r_i \in [0, r_{\max}]$ where $r_{\max} = R_\mathrm{S} \arccos(R_\mathrm{E}/R_\mathrm{S})$, corresponding to the satellite situated at the user's horizon.
Using results from trigonometry, we can calculate the Euclidean distance from the user to the satellite as a function of $r_i$: 
\begin{align} \label{equ:Euc_dis}
D(r_{i})=\sqrt{ \left( R_{h} \right)^{2}+ 4 R_\mathrm{S} R_\mathrm{E}~\sin^{2}\left( \frac{r_{i}}{2R_\mathrm{S}} \right)}.
\end{align}
Let $\alpha$ be the path loss exponent, we define the path loss as 
\begin{align} \label{equ:ell}
    \ell(r) = \left( D(r) \right)^{-\alpha}.
\end{align}

As such, let the serving satellite of the user (i.e., the nearest one) be $x_0$, we formally express the user's received SINR by
\begin{align}
    \mathrm{SINR}_0&= \frac{P_\mathrm{tx} G_0 h_0 \ell(r_{0})}{\sum_{x_i \in \Phi \cap \mathcal{B} \setminus \{x_0\}}  P_\mathrm{tx} G_i h_{i} \ell(r_i)+W} \nonumber\\
    &=\frac{h_0 \ell(r_{0})}{\sum_{x_i \in \Phi\cap \mathcal{B}  \setminus \{x_0\}} \bar{G_i} h_{i}\ell(r_i)+\bar{W}} ,
\end{align}
where $W$ represents the variance of the thermal noise, $\bar{G_i}=\frac{G_i}{G_0}<1$ and $\bar{W}=\frac{W}{P_\mathrm{tx}G_0}$ are the normalized antenna gain and noise power, respectively, and the channel fading $\{ h_i \}_{i=0}^\infty$ are i.i.d. with the following probability density function (PDF)
\begin{align}\label{equ:Nakagami-m}
    f_{h}(x) = \frac{m^m}{\Gamma(m)}x^{m-1} e^{-mx}, \quad m \in \mathbb{N}.
\end{align}

Subsequently, we define the conditional coverage probability \cite{Yang2019SINRCover} of the satellite network as
\begin{align} \label{equ:con_Ps}
    P_s(\theta)&\triangleq\mathbb{P} (\mathrm{SINR}_0 > \theta \mid \Phi, r_0 \leq r_{\max} )\nonumber\\
    &=\mathbb{P} \bigg( \frac{h_0 \ell(r_{0})}{I+\bar{W}} >\theta  \mid \Phi, r_0 \leq r_{\max} \bigg),
\end{align}
where $I \triangleq \sum_{x_i \in \Phi \cap \mathcal{B} \setminus \{x_0\}} \bar{G} h_{i} \ell(r_i)$ is the power of the accumulated interference, in which $\bar{G} = \bar{G_i}$ represents the normalized antenna gain from each interfering satellite. 

Due to the shared nature of the spectrum, simultaneous transmissions of other satellites will interfere with the user, impeding its link quality. 
In what follows, we quantify this notion by analyzing the SINR coverage probability of the considered satellite network.

\section{Analysis} \label{sec:analy}


\subsection{Regulated Spherical Point Processes}
In a similar vein to \cite{Feng2024SNC}, we define the notion of strong ball-regulated point processes on a (sky) sphere.
Concretely, let $\Phi$ be a point process generated from the probability space $(\Omega, \mathcal{F}, \mathbb{P})$ and scattered on a sphere $\mathcal{S}$ of radius $R_s$, embedded in $\mathbb{R}^3$.
We denote by $\mathcal{B}(o, r)$ a disk on the sphere (which is identical to a spherical cap), centered at an arbitrary point $o \in \mathcal{S}$, with spherical radius $r$; we denote by $\Phi(B)$ the number of points of $\Phi$ falling within a Borel set $B \subset \mathcal{S}$.
Then, we define the notion of \textit{spherical strong ball regulation} by the following.

\begin{definition}[Spherical Strong $(\sigma, \rho, \nu)$-Ball Regulation] \label{def:SSBR}
A point process $\Phi$ on $\mathcal{S}$ is strongly $(\sigma, \rho, \nu)$-ball-regulated if for all $o \in \mathcal{S}$ and $r \in [0, \pi R_s]$, the following holds
\begin{align}\label{equ:sbg}
    \Phi( \mathcal{B}(o, r) ) \le \sigma + \rho r + \nu r^2, \qquad \mathbb{P}\text{-a.s.}
\end{align}
where $\sigma, \rho, \nu$ are constants with $\sigma, \nu \ge 0$.
\end{definition}

This notion regulates the number of points within an observation window, preventing the number of nodes (or equivalently, interferers in the system model we considered) from growing unbounded. 
As such, point processes that are the strong ball-regulated exhibit locally repulsive patterns. 
For instance, a spherical hard-core point process, constructed by extending the Mat{\'e}rn Hardcore process \cite{haenggi2012HCPP} from a plane onto a sphere, fits the spherical strong ball regulation model. 
In contrast, the Poisson point process is an example that does not satisfy the strong ball-regulated model. 

Furthermore, we define the strong shot-noise-regulated point process on a sphere by the following. 
\begin{definition}[Spherical Strong $(\sigma, \rho, \nu)$-Shot-Noise Regulation] \label{def:shot-noise}
A point process $\Phi$ on $\mathcal{S}$ is strongly $(\sigma, \rho, \nu)$-shot-noise-regulated if for all non-negative, bounded, and non-increasing functions $\ell: \mathbb{R}^+ \rightarrow \mathbb{R}^+$, given spherical distances $R \in (0, \pi R_s]$, the following holds (a.s.)
\begin{align} \label{equ:shot-noise}
   \sum_{x \in \Phi \cap \mathcal{B}(o,R)} \!\!\!\!\!\!\!\! \ell(d_{\mathcal{S}} (o,x)) 
 \le \sigma \ell(0) +\! \rho \! \int_{0}^{R} \!\!\! \ell(r) dr  + 2\nu \! \int_{0}^{R} \!\!\! \ell(r) r dr.  
\end{align}
\end{definition}

In the context of a satellite communication network, Definition~\ref{def:SSBR} constrains the local density of points (the number of potential interferers), while Definition~\ref{def:shot-noise} bounds their cumulative impact (the aggregate interference). 
Moreover, we find that these two definitions are, in essence, equivalent, as formalized by the following theorem. 

\begin{theorem}
A point process $\Phi$ on $\mathcal{S}$ is strongly $(\sigma, \rho, \nu)$-shot-noise-regulated if and only if it is strongly $(\sigma, \rho, \nu)$-ball-regulated.
\end{theorem}
\begin{IEEEproof}
The proof is similar to that in \cite{Feng2024SNC} and is omitted here due to space limit.   
\end{IEEEproof}

The equivalence of these two definitions provides a powerful link between the geometric and functional properties of the point processes, enabling the derivation of a computable upper bound for the shot-noise under spherical strong ball-regulated processes.
By interpreting $\ell(r)$ as a path loss function, the shot-noise sum in \eqref{equ:shot-noise} becomes a representative model for the total interference. Consequently, the geometric condition of a strongly $(\sigma, \rho, \nu)$-ball-regulated process guarantees that the aggregated interference is bounded—a fundamental property for coverage analysis in satellite networks. 
We will leverage this principle to derive the bound on \eqref{equ:con_Ps}.

\begin{figure}[t!]
  \centering{}
  {\includegraphics[width=0.97\columnwidth]
  {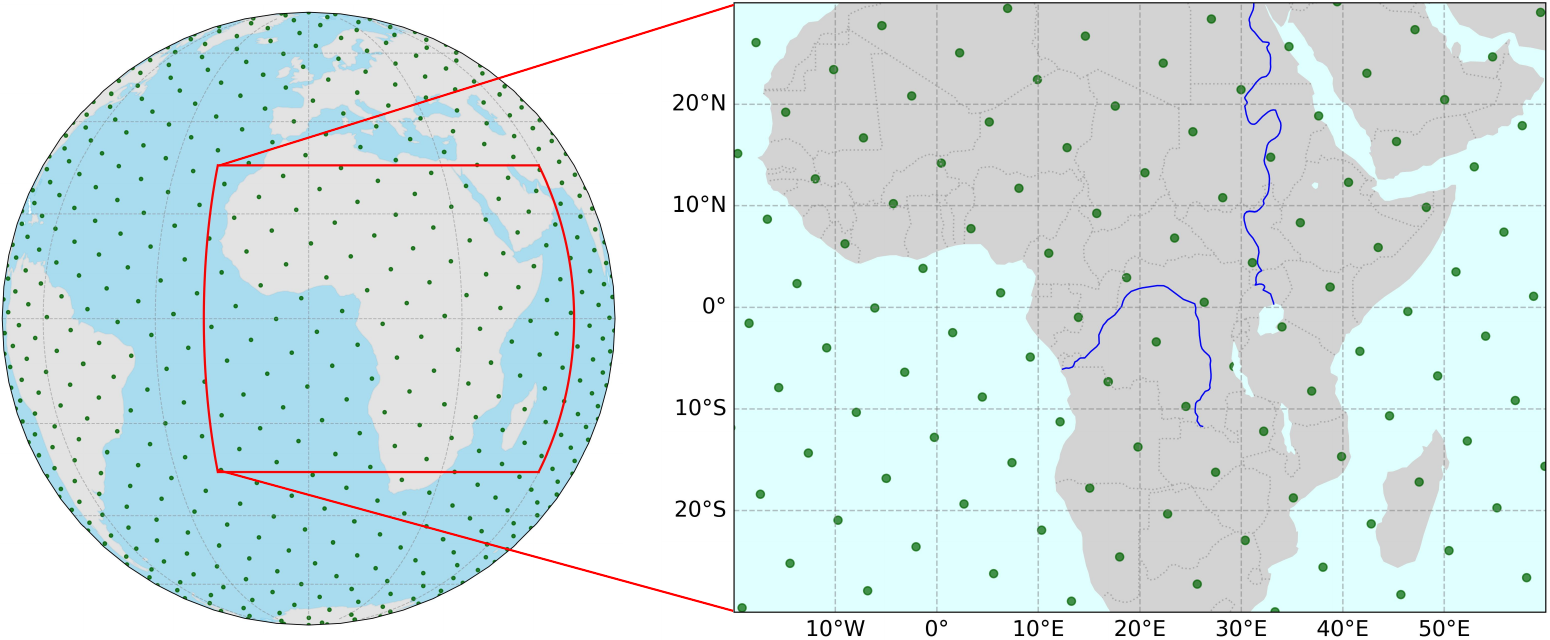}}
  \caption{Illustration of a Fibonacci lattice point distribution, showing both the global spherical view (left) and a regional projection (right).
 }\label{fig:Fibonacci}
 \vspace{-10pt}
\end{figure}

Regarding the parameter tuple $(\sigma, \rho, \nu)$, we use the Fibonacci lattice \cite{Alvaro2010Fibonacci} as a canonical example to demystify its calculation. 
Specifically, we can create a Fibonacci lattice, denoted by $\Phi_{ \mathrm{L} }$, which is comprised of $N$ (where $N$ is an odd number) points on a sphere of radius $R_s$, via the following steps. 
Let $k = \frac{N-1}{2}$, for an integer $i \in \{ -k, \dots, k\}$, the a pair of spherical coordinates $(\mathrm{lat}_i, \mathrm{lon}_i)$ (i.e., the latitude and longitude, respectively) of the $i$-th point is given by 
\begin{align}\label{equ:Fib_construed}
    \left( \mathrm{lat}_i, \mathrm{lon}_i \right) = \bigg( \arcsin\left(\frac{2i}{N}\right), \frac{ 2\pi i} {\phi} \bigg)
\end{align}
where $\phi = (1+\sqrt{5})/2$ is the golden ratio.
Moreover, denoted by $H$ half the minimum spherical distance between the nodes.
Then, according to \cite{Hardin2016ACO}, we have when $N \gg 1$, there is $2\sqrt{N}\sin(H/R_s) \approx c$ where $c$ is a constant.\footnote{When $N>200$, the approximation error falls below $0.16\%$, and can generally be considered sufficiently accurate for most practical purposes.} 
Subsequently, we establish the relationship between the number of points $N$ and the minimum distance $H$ by  
$N = (C(2\sin(H/R_s)))^2$ with the constant being $C = 3.0921$.

We display a pictorial example of the Fibonacci lattice in Fig.~\ref{fig:Fibonacci}, from which we can see that the nodes are equidistributed and quasi-uniform (such a property has been formally validated in \cite{Hardin2016ACO}). 
Therefore, we approximate the points as being uniformly distributed on both macroscopic and microscopic scales, where the density $\lambda = N/(4 \pi R_s^2)$, is determined by the ratio between the number of points $N$ and the surface area of the sphere.
Based on the approximate uniformity of the Fibonacci lattice, the number of points within a spherical cap can be bounded as follows.
\begin{proposition}
The following relationship holds for the Fibonacci lattice $\Phi_{\mathrm{L}}$:
\begin{align}
\Phi_{ \mathrm{L} }\left( \mathcal{B}(o,r) \right) \leq 1+\bigg(\frac{Cr}{4R_s\sin(H/R_s)}\bigg)^2.
\end{align}
\end{proposition}
\begin{IEEEproof}
    Considering that $\Phi_{ \mathrm{L} }\left( \mathcal{B}(o,r) \right) \le 1$, and using the approximate density $\lambda$ along with the cap area $A_s(r) = 2\pi R_s^2(1-\cos(r/R_s))$, we have
    \begin{align}
        &\Phi_{ \mathrm{L} }\left( \mathcal{B}(o,r) \right) \approx 1+\lambda A_s(r) 
        \nonumber\\
        &=1+\frac{N}{2}\left( 1 - \cos\left( \frac{r}{R_s} \right) \right) \leq 1+ \frac{N}{4 R_s^2 }r^2. 
    \end{align}
\end{IEEEproof}
This expression fits the form $\Phi_{ \mathrm{L} }\left( \mathcal{B}(o,r) \right) \le \sigma + \rho r + \nu r^2$ required by Definition~\ref{def:SSBR}. 
Therefore, we can conclude that the Fibonacci lattice is spherically strong $(\sigma, \rho, \nu)$-ball-regulated with the parameters: 
\begin{align} \label{equ:Fibon_parameter}
    ( \sigma, \rho, \nu) = \left( 1, 0, \left( \frac{3.0921}{4 R_s \sin\left( H/R_s \right)}\right)^2 \right).
\end{align}

\subsection{Coverage Analysis}
To characterize the fact that satellites in the sky do not collide with each other even though they are close by (i.e., they exhibit a locally repulsive property), we model the positions of the satellites as a spherical ball-regulated point process, with a parameter tuple $(\sigma, \rho, \nu)$. 
And the following theorem provides a (tight) bound on the conditional coverage probability.

\begin{theorem}  \label{thm:P_Nakagami}
The conditional coverage probability of the LEO satellite network is lower bounded by the following
\begin{align}
& P_s(\theta)
\ge \mathds{1}_{\{ r_0 \leq r_{\max} \}} \int_{\frac{W\theta}{\ell(r_{0})}}^{\infty} \!\!f_{h}(x)\! \Big( 1 - \exp\big(\inf_{s \in [0, s^*)} A_{\tilde{\ell}}   \nonumber\\
& \quad \quad \quad \quad \quad \quad \quad  - \tilde{\ell}(r_{0}) -\! s(x\ell(r_{0})\theta^{-1} - \bar{W}) \big) \Big)^+  dx  ,
\end{align}
where $\mathds{1}_{ \{x\} }$ is the indicator function, whilst $(x)^+ \triangleq \max(0,x)$, $s^*= m/ (\bar{G}\ell(0))$, $\tilde{\ell}(r) = -m\log(1-s\bar{G}\ell(r)/m)$, and $A_{\tilde{\ell}}$ is given as: 
\begin{align}\label{equ:A_l_tilde}
    A_{\tilde{\ell}} = \sigma\tilde{\ell}(0)+\rho \! \int_{0}^{r_{\max}}\tilde{\ell}(r)dr + 2\nu \! \int_{0}^{r_{\max}}\tilde{\ell}(r)rdr.
\end{align}
\end{theorem}

\begin{IEEEproof}
We use \eqref{equ:con_Ps} to rewrite the conditional coverage probability as the following 
   \begin{align}
    &{P}_s(\theta)=\mathbb{P} \bigg( \! I < \frac{h_0\ell(r_0)}{\theta} - \bar{W}, h_0 > \frac{\bar{W}\theta}{\ell(r_{0})} \! \mid \! \Phi,  r_0 \leq r_{\max} \! \bigg) \nonumber\\
    &= \mathds{1}_{\{ r_0 \leq r_{\max} \}}  \underbrace{\mathbb{P} \bigg( \! I < \frac{h_0\ell(r_0)}{\theta} - \bar{W}, h_0 > \frac{\bar{W}\theta}{\ell(r_{0})} \! \mid \! \Phi \! \bigg)}_{Q_1} \! .
\end{align}
Using \eqref{equ:Nakagami-m}, we can expand and bound $Q_1$ as follows:
\begin{align}
    &Q_1 
    = \int_{\frac{\bar{W}\theta}{\ell(r_{0})}}^{\infty} f_{h}(x) \Big(1 - \mathbb{P} \big(e^{sI} \ge e^{s(x\ell(r_{0})/\theta - \bar{W})}  \mid \Phi \big)\Big) dx 
    \nonumber \\ 
    &\stackrel{(a)}{\ge} \! \int_{\frac{\bar{W}\theta}{\ell(r_{0})}}^{\infty} \!\! f_{h}(x) \Big(1 -\! \! \! \inf_{s \in [0, s^*)} \! \mathcal{L}_{I \mid \Phi}(-s) e^{-s(x\ell(r_{0})/\theta - \bar{W})}\Big)^+ \! dx 
    \nonumber \\ 
    &\stackrel{(b)}{\ge} \! \int_{\frac{\bar{W}\theta}{\ell(r_{0})}}^{\infty} \!\! f_{h}(x) \Big( 1 -  e^{\inf_{s \in [0, s^*)}  A_{\tilde{\ell}} - \tilde{\ell}(r_{0})- s(\frac{x\ell(r_{0})}{\theta} - \bar{W}) } \Big)^+ \!   dx,
   \end{align} 
   where ($a$) follows from the Cheffnoff bound, i.e., $\mathbb{P}(I>x\mid \Phi) \le \exp(-sx)\mathcal{L}_{I\mid \Phi}(-s)$ and ($b$) holds due to the following
   \begin{align}
       \mathcal{L}_{I\mid \Phi }(-s) 
       &=\exp\left( \sum_{x_i\in \Phi \cap \mathcal{B} \setminus \{x_0\}} \!\!\!\!\!\! \log \mathcal{L}_{h}\Big( \! -s \bar{G} \ell(r_i) \Big) \right) \nonumber\\
       & \stackrel{(c)}{\le} \exp(A_
       {\tilde{\ell}})-\tilde{\ell}(r_0),
   \end{align}
   in which we let $\tilde{\ell}(r)=\log \mathcal{L}_h(-s \bar{G}\ell(r))$ and ($c$) results from noticing that $\mathcal{L}_h(s)=(1+s/m)^{-m}$ and then applying the shot-noise regulation definition \eqref{equ:shot-noise}. 

   The ultimate result follows from algebraic simplifications to the terms.
\end{IEEEproof}

Theorem~\ref{thm:P_Nakagami} establishes a general lower bound on the link performance for channels subject to Nakagami-\textit{m} fading with any parameter $m \ge 0.5$. For the special case of Rayleigh fading ($m=1$), this bound simplifies significantly, as stated in the following corollary.

\begin{corollary}\label{cor:P_Rayleigh}
Under Rayleigh fading, the conditional coverage probability of the LEO satellite network can be lower bounded as 
\begin{align}
    P_s(\theta) \ge \mathds{1}_{\{ r_0 \leq r_{\max} \}} \exp\left( \tilde{\ell}(r_{0}) -\frac{\theta \bar{W}}{\ell(r_{0})}  -A_{\tilde{\ell}}  \right),
\end{align}
where $\tilde{\ell}(r) =\log\big( 1+\theta \bar{G} \ell(r)/\ell(r_{0}) \big)$ and $A_{\tilde{\ell}}$ is given in \eqref{equ:A_l_tilde}.
\end{corollary}

\begin{IEEEproof}
When the channel fading follows Rayleigh distribution, the conditional coverage probability can be calculated as
    \begin{align} \label{equ:con_P_Rayleigh}
     &P_s(\theta)  
     = \mathds{1}_{\{ r_0 \leq r_{\max} \}} e^{ -\frac{\theta \bar{W}}{\ell(r_{0})} }  \!\!\! \prod_{x_i\in \Phi \cap \mathcal{B}  \backslash\{x_0\}} \frac{1}{1+\theta \bar{G} \frac{\ell(r_i)}{\ell(r_{0})}} \nonumber \\
     &= \mathds{1}_{\{ r_0 \leq r_{\max} \}} \cdot \nonumber \\
     & \quad \exp \bigg( \! -\frac{\theta \bar{W}}{\ell(r_{0})} - \!\!\!  \sum_{x_i\in \Phi \cap \mathcal{B} \backslash \{x_0\} } \!\!\!\! \log \bigg(1+\theta \bar{G} \frac{\ell(r_i)}{\ell(r_{0})} \bigg)\Bigg).
    \end{align}
    Applying \eqref{equ:shot-noise} with $r_{\max}$ to the total interference provides an upper bound, which in turn yields the coverage probability lower bound.
\end{IEEEproof}

Regarding the application of the above theoretical results, we take the Fibonacci lattice as an example.
Suppose the satellites are deployed according to this point process (cf. Fig.~\ref{fig:Fibonacci}), we can calculate the $(\sigma, \rho, \nu)$ pair as per \eqref{equ:Fibon_parameter}. 
By substituting the values of \eqref{equ:Fibon_parameter} into \eqref{equ:A_l_tilde}, we obtain $A_{\tilde{\ell}}$. This subsequently yields the lower bound for the coverage probability under Nakagami-\textit{m} fading and Rayleigh fading.

A key advantage of our derived bounds is their computational efficiency. To quantify this, we evaluated the running times for Theorem~\ref{thm:P_Nakagami} and Corollary~\ref{cor:P_Rayleigh} using MATLAB 2024b (with $m=1$ and $\theta$ spanning -15 to 15 in steps of 1). The computation took 37.093 seconds for Theorem~\ref{thm:P_Nakagami} and 0.033 seconds for Corollary~\ref{cor:P_Rayleigh}. 
This comparison also highlights the significant computational gain provided by the more concise expression in Corollary~\ref{cor:P_Rayleigh}.

\section{Simulation and Numerical Results}
In this section, we validate the effectiveness of our analytical framework by comparing the theoretical bounds with Monte Carlo simulation results obtained from Starlink constellations.
For the analysis, we model the satellite constellation as a Fibonacci lattice on the satellite sphere (with radius $R_\mathrm{S}$), constructed according to \eqref{equ:Fib_construed}. The user is located at the North Pole of the Earth (where the radius is $R_\mathrm{E}=6370$ km). 
Note that the serving satellite can be precisely identified for this configuration. 
Specifically, within the construction defined by \eqref{equ:Fib_construed}, the point corresponding to $i=k$ is the closest to the user. 
And the distance $r_0$ is thus given by $r_0=R_\mathrm{S} \arccos((N-1)/N)$.
Correspondingly, the specific regulation parameters $(\sigma, \rho, \nu)$ used to evaluate the theoretical lower bound for this lattice are given by \eqref{equ:Fibon_parameter}.
In order to compare the performance evaluated from our theoretical framework to that of an actual satellite deployment, we choose the Starlink constellation [xx] as our reference. 
We obtained the Starlink topology data from CelesTrak for satellites with an altitude of $R_\mathrm{h}=550$ km. Specifically, the simulation uses a snapshot of satellite locations visible from Beijing, China, at a specific time instance.
Unless otherwise specified, we take the path loss exponent $\alpha = 2$ \cite{Wang2022UldenseLEO}, the normalized noise power $\bar{W} = 0$, the orbital altitude $R_\mathrm{h}=500$ km, and the gain ratio of the antennas of the interference satellite and serving satellite $\bar{G}= 0.1$. 
The user has a minimum elevation angle of $\omega_{\min}=25\degree$, and the channel is characterized by a Nakagami-\textit{m} fading parameter of $m=2$.
The statistics are collected by averaging through  $5\times 10^4$ simulation runs.

\begin{figure}[t!]
  \centering{}
    {\includegraphics[width=0.97\columnwidth]{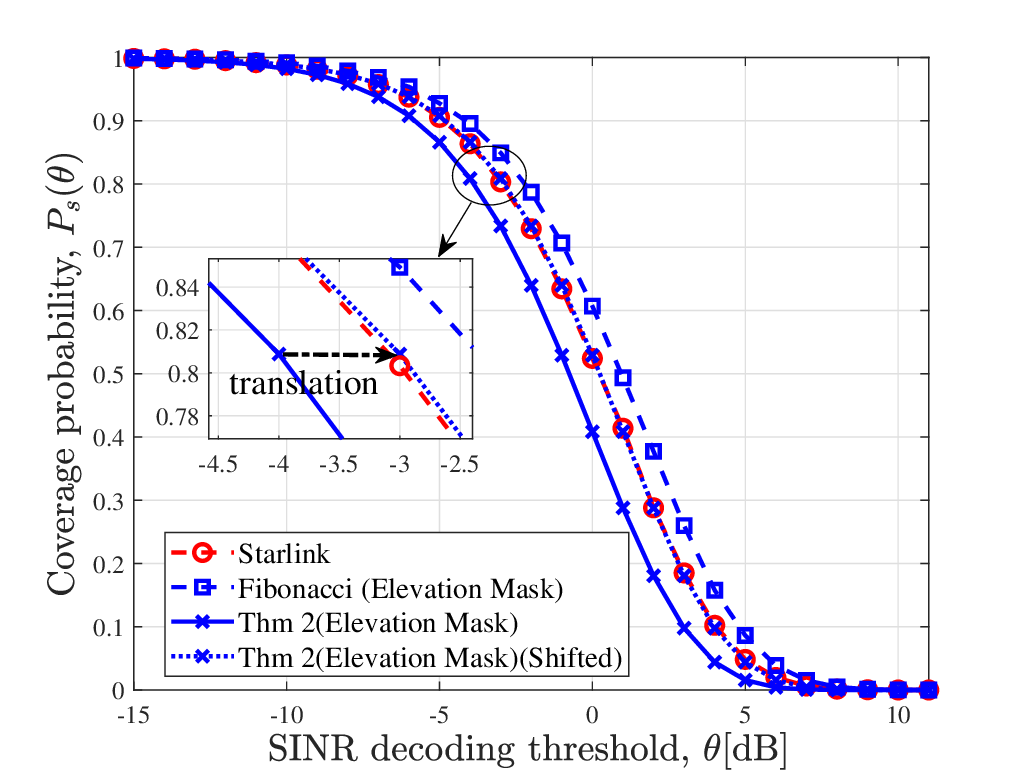}}
  \caption{Coverage probability under simulations and analysis, with the number of visible satellites $N_{vis}=16$, a minimum elevation angle $\omega_{\min}=25\degree$, and a Nakagami-\textit{m} fading parameter of $m=2$.
 }
  \label{fig:appro_real}
  \vspace{-10pt}
\end{figure}

Fig.~\ref{fig:appro_real} displays the coverage probability obtained under simulations and analysis.
To ensure fairness, the average number of visible satellites (denoted by $N_{vis}$) is kept constant (as $N_{vis}=16$) for both the Starlink deployment and the corresponding Fibonacci lattice model (with a total of $N=2973$ satellites, where $r_0=179.46$ km and $H=196.2013$ km). 
The plot yields two key insights. First, between the two simulation curves, the Fibonacci lattice results in a slightly better coverage performance than the Starlink snapshot. 
This is attributable to the lattice's more uniform point distribution, which reduces the probability of encountering severe interference compared to the practical deployment. 
Second, this figure validates that Theorem~\ref{thm:P_Nakagami} acts as a tight lower bound for the Fibonacci lattice (which is its intended purpose).
In addition, we make a significant observation regarding the Starlink simulation. Although the Starlink data represents only a single snapshot used to illustrate a performance trend, the shape of its curve is remarkably similar to that given by Theorem~\ref{thm:P_Nakagami}. 
As shown in Fig.~\ref{fig:appro_real}, a simple horizontal shift of the Theorem~\ref{thm:P_Nakagami} curve by approximately 1 dB results in a near-perfect overlay with the Starlink simulation. 
While this is an observation of a specific instance and not a formal proof of equivalence, it strongly suggests that our analytical framework accurately captures the fundamental performance trend dictated by spatial constraints. This indicates the potential for our framework to be adapted as a reliable performance estimator for real-world networks.

\begin{figure}[t!]
  \centering{}
    {\includegraphics[width=0.97\columnwidth]{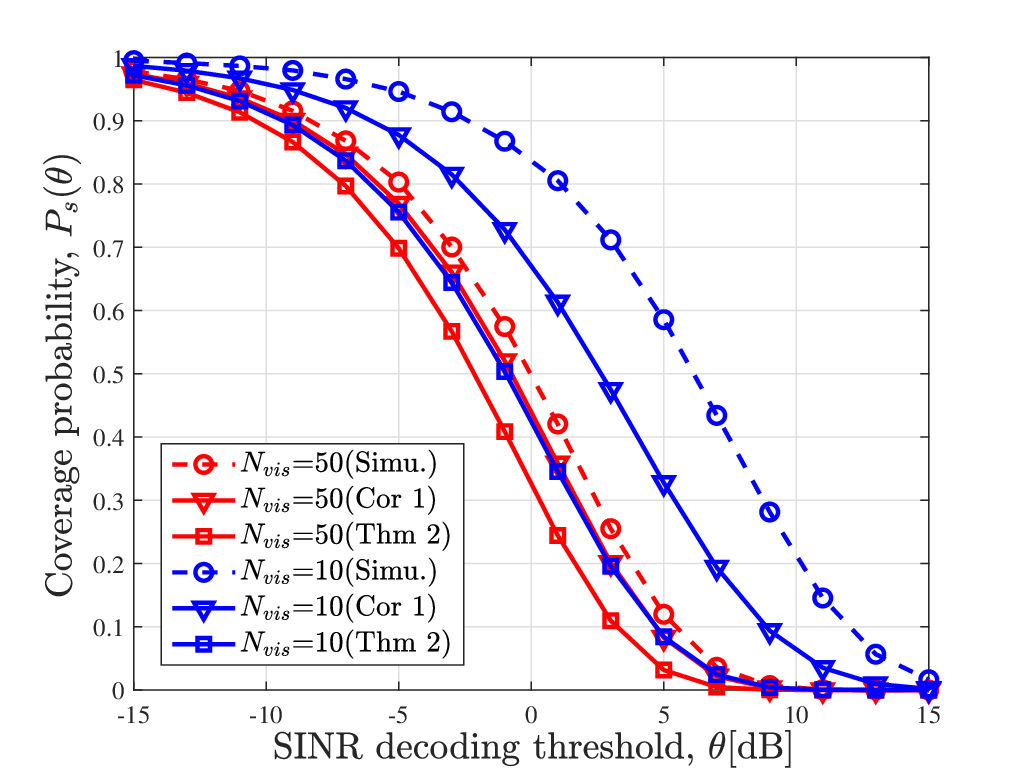}}
  \caption{Coverage probability versus the decoding threshold, $\theta$, comparing the analytical results of Theorem~\ref{thm:P_Nakagami} (with $m=1$) and Corollary~\ref{cor:P_Rayleigh} with simulations for two density scenarios: $N_{vis}=10$ (with $H=650$ km) and $N_{vis}=50$ (with $H=287$ km).
 }
  \label{fig:Rayleigh&density}
  \vspace{-10pt}
\end{figure}


Fig.~\ref{fig:Rayleigh&density} validates the analytical bounds from Corollary~\ref{cor:P_Rayleigh} and Theorem~\ref{thm:P_Nakagami} against simulations of the Fibonacci lattice model, performed for different satellite densities under a Rayleigh fading channel. The plot shows the coverage probability as a function of the decoding threshold, $\theta$. 
The two density scenarios are represented by the number of visible satellites.
The high-density case ($N_{vis}=50$) corresponds to a small $H$ (half the minimum spherical distance) of $287$ km, $N=1369$, and $r_0=262.60$ km. The low-density case ($N_{vis}=10$) corresponds to a large $H$ of $650$ km, $N=267$, and $r_0=594.77$ km.
In both cases, the analytical results are shown to tightly bound the simulation curves. Notably, the bound derived from Corollary~\ref{cor:P_Rayleigh} is consistently tighter than that from Theorem~\ref{thm:P_Nakagami}.

Increasing the satellite density has a twofold effect. On the one hand, a higher density leads to stronger co-channel interference, which degrades the overall link performance. On the other hand, it also causes the point distribution to become more tightly packed and uniform. This increased regularity allows the assumptions underpinning our theoretical analysis to hold more accurately, which in turn improves the effectiveness of our derived lower bound. 
Therefore, while the analytical and simulation curves show strong agreement in their trends for both density cases, the gap between them is notably different, being smaller in the high-density scenario.

\begin{figure}[t!]
  \centering{}
    {\includegraphics[width=1\columnwidth]{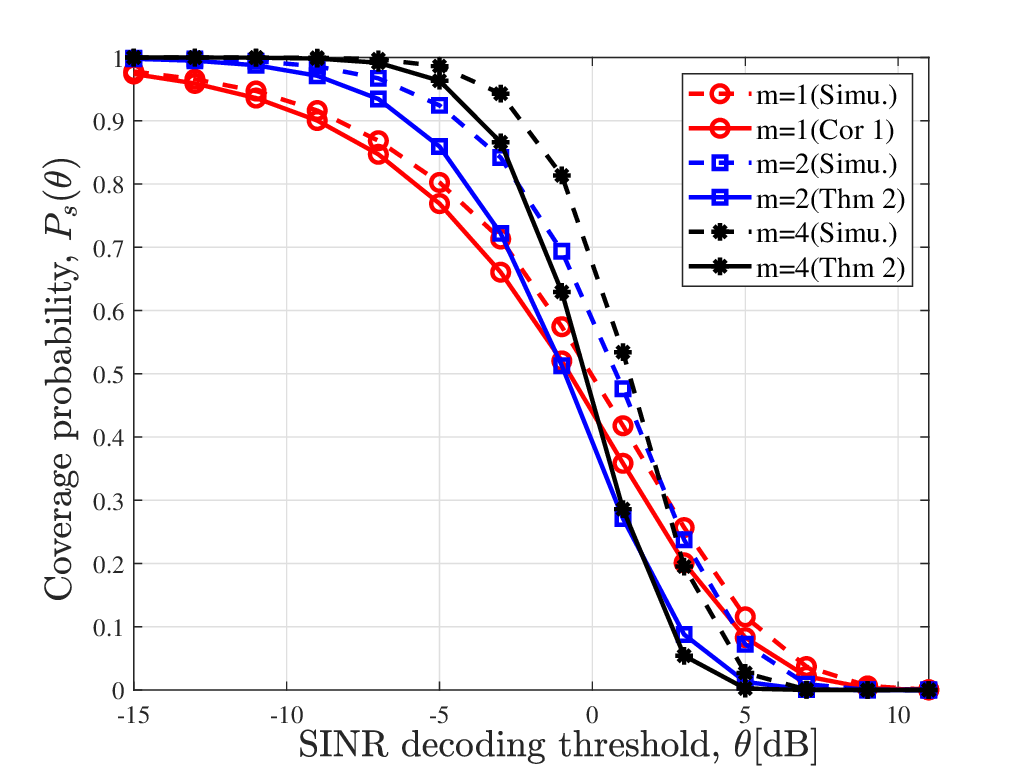}}
  \caption{The coverage probability for various Nakagami-\textit{m} fading parameters ($m \in \{1, 2, 4\}$) and a fixed number of visible satellites of $N_{vis}=50$.
 }
  \label{fig:comp_channel_mode}
  \vspace{-10pt}
\end{figure}

The Nakagami-\textit{m} fading model provides a more general and often more realistic representation of satellite channels compared to Rayleigh fading. Fig.~\ref{fig:comp_channel_mode} therefore evaluates the performance of our analytical bounds under different small-scale fading conditions by varying the Nakagami-\textit{m} parameter, $m$. It is observed that the analytical framework remains tight under different fading conditions, providing robust and effective bounds.

\section{Conclusion}
In this paper, we developed a theoretical framework for the performance analysis of LEO satellite networks, substantially improving the modeling accuracy and concurrently reducing computational complexity.
Specifically, we leveraged the spatial network calculus to define a set of spherical strong ball-regulated point processes, which we used to model the satellites' spatial positions. 
Such point processes exhibit a locally repulsive pattern, reflecting the fact that nearby satellites are separated by a safety distance and do not collide. 
Subsequently, we derived analytical expressions for the lower bounds of the conditional coverage probability under Nakagami-\textit{m} and Rayleigh fading, respectively. 
The expressions have a low computational complexity, enabling efficient numerical evaluations. 
We validated the accuracy of our mathematical derivations by contrasting the analysis with the coverage probability estimated from a Starlink constellation. 
The numerical results showed that the analysis provides a tight lower bound to the simulations, and surprisingly, they aligned almost perfectly through a 1 dB shift. 
This corroborated the efficacy of our model, demonstrating that it serves as an accurate theoretical model for LEO satellite networks.


\bibliographystyle{IEEEtran}
\bibliography{bib/Analysis_of_NTN_by_SNC}

\end{document}

%% file: SupportDocuments/aconym.tex
\acrodef{CCDF}{complementary cumulative distribution function}
\acrodef{CF}{characteristic function}
\acrodef{PPP}{Poisson point processe}
\acrodef{RV}{random variable}
\acrodef{i.i.d.}{independent and identically distributed}
\acrodef{PDF}{probability distribution function}
\acrodef{CDF}{cumulative distribution function}
\acrodef{ch.f.}{characteristic function}
\acrodef{AWGN}{additive white Gaussian noise}
\acrodef{SNR}{signal-to-noise ratio}
\acrodef{LRT}{likelihood ratio test}
\acrodef{DRT}{distance ratio test}
\acrodef{GLRT}{generalized likelihood ratio test}
\acrodef{CRLB}{Cram\'{e}r-Rao lower bound}
\acrodef{CRB}{Cram\'{e}r-Rao bound}
\acrodef{ZZLB}{Ziv-Zakai lower bound}
\acrodef{ZZB}{Ziv-Zakai bound}
\acrodef{LOS}{line-of-sight}
\acrodef{ToF}{time-of-flight}
\acrodef{NLOS}{non-line-of-sight}
\acrodef{GDOP}{geometric dilution of precision}
\acrodef{GPS}{Global Positioning System}
\acrodef{FIM}{Fisher information matrix}
\acrodef{PEB}{position error bound}
\acrodef{SPEB}{squared position error bound}
\acrodef{TOA}{time-of-arrival}
\acrodef{TOF}{time-of-flight}
\acrodef{WSN}{wireless sensor network}
\acrodef{MAC}{medium access control}
\acrodef{RSS}{received signal strength}
\acrodef{WAF}{wall attenuation factor}
\acrodef{TDOA}{time difference-of-arrival}
\acrodef{RF}{radiofrequency}
\acrodef{RTT}{round-trip time}
\acrodef{AOA}{angle-of-arrival}
\acrodef{MF}{matched filter}
\acrodef{ED}{energy detector}
\acrodef{ML}{maximum likelihood}
\acrodef{MSE}{mean-square error}
\acrodef{RMSE}{root-mean-square error}
\acrodef{LEO}{localization error outage}
\acrodef{ppm}{part-per-million}
\acrodef{ACK}{acknowledge}
\acrodef{UWB}{Ultrawide bandwidth}
\acrodef{TNR}{threshold-to-noise ratio}
\acrodef{LS}{least squares}
\acrodef{IR-UWB}{impulse radio UWB}
\acrodef{FCC}{Federal Communications Commission}
\acrodef{TH}{time-hopping}
\acrodef{PPM}{pulse position modulation}
\acrodef{MUI}{multi-user interference}
\acrodef{PDP}{power delay profile}
\acrodef{BPZF}{band-pass zonal filter}
\acrodef{SIR}{signal-to-interference ratio}
\acrodef{SINR}{signal-to-interference-plus-noise ratio}
\acrodef{RFID}{radio frequency identification}
\acrodef{WPAN}{wireless personal area network}
\acrodef{WWB}{Weiss-Weinstein bound}
\acrodef{DP}{direct path}
\acrodef{MF}{matched filter}
\acrodef{MMSE}{minimum-mean-square-error}
\acrodef{SBS}{serial backward search}
\acrodef{SBSMC}{serial backward search for multiple clusters}
\acrodef{NBI}{narrowband interference}
\acrodef{WBI}{wideband interference}
\acrodef{INR}{interference-to-noise ratio}
\acrodef{CR}{channel response}
\acrodef{CIR}{channel impulse response}
\acrodef{CR}{channel  response}
\acrodef{RADAR}{radar}
\acrodef{MUR}{Multistatic radar}
\acrodef{JBSF}{jump back and search forward}
\acrodef{HDSA}{high-definition situation-aware}
\acrodef{RRC}{root raised cosine}
\acrodef{ST}{simple thresholding}
\acrodef{BTB}{Bellini-Tartara bound}
\acrodef{P-Max}{$P$-Max}  
\acrodef{MIMO}{multiple-input multiple-output}
\acrodef{MAP}{maximum a posteriori}
\acrodef{FG}{factor graph}
\acrodef{OP}{outage probability}
\acrodef{WED}{wall extra delay}
\acrodef{RMS}{root mean square}
\acrodef{SPAWN}{sum-product algorithm over a wireless network}
\acrodef{MDD}{minimum distance distribution}
\acrodef{MAP}{maximum a posteriori probability}
\acrodef{SAP}{small cell access point}
\acrodef{UE}{user equipment}
\acrodef{MBS}{macro cell base station}
\acrodef{UER}{\ac{UE} Relay}
\acrodef{D2D}{device-to-device}
\acrodef{MBS}{macro base station}
\acrodef{CSI}{channel state information}
\acrodef{OGR}{outage guard region}
\acrodef{FUR}{feasible UER region}
\acrodef{EHR}{energy harvesting region}
\acrodef{EH}{energy harvesting}
\acrodef{D2D-EHSN}{D2D communication provided \ac{EH} small cell network}
\acrodef{D2D-EHHN}{D2D communication provided \ac{EH} heterogeneous network}
\acrodef{3GPP}{3rd Generation Partnership Project}
\acrodef{BS}{base station}
\acrodef{DF}{decode and forward}
\acrodef{CCDF}{complementary cumulative distribution function}
\acrodef{ZF}{zero forcing}
\acrodef{RZF}{regularized zero forcing}
\acrodef{WLLN}{weak law of large number}
\acrodef{SLLN}{strong law of large numbers}
\acrodef{TDD}{Time-division duplex}
\acrodef{EE}{energy efficiency} 
\acrodef{HetNet}{heterogeneous network} 
\acrodef{SCP}{Single Cell Processing}
\acrodef{CBF}{Coordinated Beamforming}

%% file: SupportDocuments/defmetric.tex
\usepackage{color}
\usepackage{dsfont}
\usepackage{bbm}

\DeclareMathAlphabet{\mathsf}{OML}{cmbr}{m}{it}

\newtheorem{definition}{\bf Definition}
\newtheorem{theorem}{\bf Theorem}

\newtheorem{corollary}{\bf Corollary}
\newtheorem{proposition}{\bf Proposition}

\newcommand{\bd}{\begin{description}}
\newcommand{\ed}{\end{description}}
\newcommand{\be}{\begin{enumerate}}
\newcommand{\ee}{\end{enumerate}}
\newcommand{\bi}{\begin{itemize}}
\newcommand{\ei}{\end{itemize}}
\newcommand{\bl}{\begin{list}}
\newcommand{\el}{\end{list}}
\newcommand{\bt}{\begin{tabbing}}
\newcommand{\et}{\end{tabbing}}

%% file: bib/Analysis_of_NTN_by_SNC.bib
@STRING{IEEE_J_WCOML = "IEEE Wireless Commun. Lett."}

@STRING{IEEE_J_TWC = "IEEE Trans. Wireless Commun."}

@STRING{IEEE_J_TCOM = "IEEE Trans. Commun."}

@STRING{IEEE_J_JSAC = "IEEE J. Sel. Areas Commun."}

@STRING{IEEE_J_TIT = "IEEE Trans. Inf. Theory"}

@book{haenggi2012HCPP,
  title={Stochastic Geometry for Wireless Networks},
  author={Haenggi, Martin},
  year={2012},
  publisher={Cambridge University Press, Cambridge}
}

@article{WanKisYan:25TAES,
  title={ Analyzing Localizability of {LEO/MEO} Hybrid Networks: A Stochastic Geometry Approach },
  author={ Wang, R. and Kishk, M. A. and Yang, H.~H. and Alouini, M.-S.},
  journal={IEEE Trans. Aerosp. Electron. Syst.},
  volume={61},
  number={4},
  pages={10720--10736}, 
  year = {Aug. 2025}
}

@ARTICLE{Zhang2025PAHPP,
  author={Zhang, Haoxing and Miao, Xiaqing and Ni, Zihan and Wang, Shuai and Pan, Gaofeng and Cavdar, Cicek and An, Jianping},
  journal=IEEE_J_TWC, 
  title={{LEO} Mega-Constellation-Terrestrial Communications Suffering Poisson Arc Hardcore Distributed Space Interference}, 
  year={Apr. 2025},
  volume={24},
  number={4},
  pages={2707-2721}}

@ARTICLE{Zhong2025SNC,
  author={Zhong, Yi and Zhou, Xiaohang and Feng, Ke},
  journal=IEEE_J_TWC, 
  title={{Spatial Network Calculus: Toward Deterministic Wireless Networking}}, 
  note = {accepted to appear},
  year={2025}
}

@ARTICLE{Feng2024SNC,
  author={Feng, Ke and Baccelli, François},
  journal=IEEE_J_TWC, 
  title={Spatial Network Calculus and Performance Guarantees in Wireless Networks}, 
  year={May. 2024},
  volume={23},
  number={5},
  pages={5033-5047}
  }

@article{Alvaro2010Fibonacci,
author = {Gonz{\'a}lez, {\'A}lvaro},
year = {Jan. 2010},
pages = {49-64},
title = {Measurement of Areas on a Sphere Using Fibonacci and Latitude–Longitude Lattices},
volume = {42},
journal = {Math. Geosci.}
}

@article{Hardin2016ACO,
  title={A Comparison of Popular Point Configurations on $\mathbb{S}^2$},
  author={Douglas P. Hardin and Timothy Michaels and Edward B. Saff},
  journal={Dolomites Res. Notes Approx.},
  year={Jul. 2016},
  volume={9}
}

@article{Park2023ATA,
  title={A Tractable Approach to Coverage Analysis in Downlink Satellite Networks},
  author={Jeonghun Park and Jinseok Choi and Namyoon Lee},
  journal=IEEE_J_TWC,
  year={Feb. 2023},
  volume={22},
  pages={793-807}
  }

@ARTICLE{AlHourani2021AnAA,
  author={Al-Hourani, Akram},
  journal=IEEE_J_WCOML, 
  title={An Analytic Approach for Modeling the Coverage Performance of Dense Satellite Networks}, 
  year={Apr. 2021},
  volume={10},
  number={4},
  pages={897-901}}

@article{Okati2021NonhomogeneousSG,
  title={Nonhomogeneous Stochastic Geometry Analysis of Massive {LEO} Communication Constellations},
  author={Niloofar Okati and Taneli Riihonen},
  journal=IEEE_J_TCOM,
  year={Mar. 2022},
  volume={70},
  number={3},
  pages={1848-1860}
}

@article{Okati2020DownlinkCA,
      title={Downlink Coverage and Rate Analysis of Low Earth Orbit Satellite Constellations Using Stochastic Geometry},
      author={Niloofar Okati and Taneli Riihonen and Dani Korpi and Ilari Angervuori and Risto Wichman},
      journal=IEEE_J_TCOM,
      year={Aug. 2020},
      volume={68},
      pages={5120-5134}
      }

@misc{kim2024spectrumsharing,
  title={Spectrum Sharing Between Low Earth Orbit Satellite and Terrestrial Networks: A Stochastic Geometry Perspective Analysis}, 
  author={Daeun Kim and Jeonghun Park and Jinseok Choi and Namyoon Lee},
  year={2024},
  eprint={2408.12145},
  archivePrefix={arXiv},
  primaryClass={eess.SP},
  url={https://arxiv.org/abs/2408.12145}, 
}

@article{Kim2023CoverageAO,
  title={Coverage Analysis of Dynamic Coordinated Beamforming for {LEO} Satellite Downlink Networks},
  author={Daeun Kim and Jeonghun Park and Namyoon Lee},
  journal=IEEE_J_TWC,
  year={Sep. 2023},
  volume={23},
  pages={12239-12255}
}

@ARTICLE{Yang2019SINRCover,
  author={Yang, Howard H. and Quek, Tony Q. S.},
  journal=IEEE_J_TCOM, 
  title={Spatio-Temporal Analysis for {SINR} Coverage in Small Cell Networks}, 
  year={May. 2019},
  volume={67},
  number={8},
  pages={5520-5531}}

@ARTICLE{Talgat2021StocGA,
  author={Talgat, Anna and Kishk, Mustafa A. and Alouini, Mohamed-Slim},
  journal="IEEE Commun. Lett.", 
  title={Stochastic Geometry-Based Analysis of {LEO} Satellite Communication Systems}, 
  year={Aug. 2021},
  volume={25},
  number={8},
  pages={2458-2462}}

@ARTICLE{Wang2022UldenseLEO,
  author={Wang, Ruibo and Kishk, Mustafa A. and Alouini, Mohamed-Slim},
  journal="IEEE Commun. Mag", 
  title={Ultra-Dense {LEO} Satellite-Based Communication Systems: A Novel Modeling Technique}, 
  year={Apr. 2022},
  volume={60},
  number={4},
  pages={25-31}}

@inproceedings{LuYanPap:23Globecom,
  title={ Analysis of Age of Information in Non-terrestrial Networks },
  author={Lu, Y. and Yang, H. H. and Pappas, N. and Geraci, G. and Quek, T. Q. S. },
  booktitle={Proc. IEEE Global Commun. Conf. (Globecom) Workshop},
  year = {Dec. 2023}, 
  Address = {Kuala Lumpur, Malaysia}
}

@ARTICLE{Haenggi2009steo,
  author={Haenggi, Martin and Andrews, Jeffrey G. and Baccelli, Francois and Dousse, Olivier and Franceschetti, Massimo},
  journal=IEEE_J_JSAC, 
  title={Stochastic geometry and random graphs for the analysis and design of wireless networks}, 
  year={Sep. 2009},
  volume={27},
  number={7},
  pages={1029-1046}}

@ARTICLE{Andrews2011Coverage,
  author={Andrews, Jeffrey G. and Baccelli, Francois and Ganti, Radha Krishna},
  journal=IEEE_J_TCOM, 
  title={A Tractable Approach to Coverage and Rate in Cellular Networks}, 
  year={Nov. 2011},
  volume={59},
  number={11},
  pages={3122-3134}}

@ARTICLE{Haenggi2013delay,
  author={Haenggi, Martin},
  journal=IEEE_J_TIT, 
  title={The Local Delay in Poisson Networks}, 
  year={Mar. 2013},
  volume={59},
  number={3},
  pages={1788-1802}}

@ARTICLE{Zhang2015Throughput,
  author={Zhang, Xinchen and Andrews, Jeffrey G.},
  journal=IEEE_J_TCOM, 
  title={Downlink Cellular Network Analysis With Multi-Slope Path Loss Models}, 
  year={May. 2015},
  volume={63},
  number={5},
  pages={1881-1894}}
